\documentclass[a4paper,twoside]{article}

\usepackage{graphicx}
\usepackage{subcaption}
\usepackage{todonotes}
\usepackage{booktabs}  
\usepackage{cite}
\usepackage{float}
\usepackage{makecell}
\usepackage{lipsum}

\usepackage{epsfig}
\usepackage{subcaption}
\usepackage{calc}
\usepackage{amssymb}
\usepackage{amstext}
\usepackage{amsmath}
\usepackage{amsthm}
\usepackage{multicol}
\usepackage{pslatex}
\usepackage{apalike}
\usepackage{caption}
\DeclareCaptionType{mycapequ}[][List of equations]
\UseRawInputEncoding
\usepackage[bottom]{footmisc}
\usepackage{SCITEPRESS}     % Please add other packages that you may need BEFORE the SCITEPRESS.sty package.

\begin{document}

% \title{Joint Segmentation of Layers and Fluids in Retinal OCT Images Using Deep Learning}
\title{Retinal Image Segmentation with Small Datasets}

%Authors and affiliations are anonymised for double blind checking by Bioimaging.
% \author{\authorname{Nchongmaje Ndipenoch\sup{1}\orcidAuthor{0000-0000-0000-0000}, Yongmin Li\sup{2}\orcidAuthor{0000-0000-0000-0000} and Alina Miron \sup{2}\orcidAuthor{0000-0000-0000-0000}}
% \affiliation{\sup{1,2}Department of Computer Science, Brunel University London, Uxbridge, UB8 3PH, United Kingdom}}

\author{\authorname{Nchongmaje Ndipenoch, Alina Miron, Zidong Wang and Yongmin Li}
\affiliation{Department of Computer Science, Brunel University London, Uxbridge, UB8 3PH, United Kingdom}}

%\email{\{f\_author, s\_author\}@ips.xyz.edu, t\_author@dc.mu.edu}


\abstract{Many eye diseases like Diabetic Macular Edema (DME), Age-related Macular Degeneration (AMD), and Glaucoma manifest in the retina,  can cause irreversible blindness or severely impair the central version. The Optical Coherence Tomography (OCT), a 3D scan of the retina with high qualitative information about the retinal morphology, can be used to diagnose and monitor changes in the retinal anatomy. Many Deep Learning (DL) methods have shared the success of developing an automated tool to monitor pathological changes in the retina. However, the success of these methods depend mainly on large datasets. 
% In a domain where obtaining a dataset is highly challenging and often very small and limited, 
To address the challenge from very small and limited datasets,
we proposed a DL architecture termed CoNet (Coherent Network) for joint segmentation of layers and fluids in retinal OCT images on very small datasets (less than a hundred training samples). The proposed model was evaluated on the publicly available Duke DME dataset consisting of 110 B-Scans from 10 patients suffering from DME.
Experimental results show that the proposed model outperformed both the human experts' annotation and the current state-of-the-art architectures by a clear margin with a mean Dice Score of 88\% when trained on 55 images without any data augmentation.
}

\onecolumn \maketitle \normalsize \setcounter{footnote}{0} \vfill

\section{\uppercase{Introduction}}
\label{sec:introduction}

Diabetic retinopathy (DR), 
a disease that damages the blood vessels in the retina,
is the most common cause of blindness among working-aged adults in the United States \cite{kles:2007}. Among those affected, approximately 21 million people develop diabetic macular edema (DME) \cite{bresnick:Ophthalmology1986}. 
% DME is an eye disease which can occur in diabetic patients. 
DME is the accumulation of fluid in the macula that can damage the blood vessels in the eye due to high blood sugar over time. The macula is the retina's centre at the back of the eye, where vision is the sharpest. 
% The disease that damages the blood vessels in the retina is called Diabetic retinopathy.

Presently an effective treatment of eye diseases exists in the form of anti-vascular endothelial growth factor (anti-VEGF) therapy \cite{shienbaum:AJO2013}. However, the effectiveness of the treatment depends on early diagnosis and frequent monitoring of the progress of the disease. Also, anti-VEGF drugs are expensive and need to be administered regularly.

Early diagnosis, effective frequent monitoring, and behavioural advice from ophthalmologists, such as diets and regular exercises, are key factors in preventing or slowing down the disease's progress. Still, as of today, these are mostly done manually, which is time-consuming, laborious, and prone to errors. Hence, there is the need to develop an automated tool to monitor retinal morphology and fluid accumulation properly. 

The Optical Coherence Tomography (OCT), a high-resolution 3D non-invasive imaging modality of the retina acquiring a series of cross-sectional slices (B-scans), provides qualitative information and visualisations of the retinal structure. The development of an automated method to study the retina anatomy from OCT B-Scans and hence the evaluation of eye pathogens like DME will be of high value and importance.

% To help resolve the above problem, 
To address the above problem, 
we propose a deep learning based model,
termed CoNet (Coherent Network),
for simultaneously segmenting layers and fluid in retinal OCT B-Scans. In contrast to the common approach of treating retinal layers and fluid regions separately, CoNet provides an automatic solution for simultaneously segmenting both.

The rest of the paper is organized as follows. A brief review of the previous studies is provided in Section~\ref{sec:background}. The description of the proposed method is put forward in Section~\ref{sec:methods}. The  experiments and result analysis are presented in Section~\ref{sec:experiments}. Finally, the conclusion with our contributions is described in Section~\ref{sec:conclusion}.

\section{\uppercase{Background}}
\label{sec:background}

The OCT was developed in the 1990s by \cite{Huang:science1991} but only became commercially available in 2006. It permits fast image acquisition and success in quantitative analysis because of its high quality and resolution. Some of the earliest segmentation approaches of retinal images include: 
segmentation of retinal layers in OCT images using the graph method \cite{garvin:IEEETMI2009}, segmentation of fluid in the retina in patients suffering from Macular Edema (ME) by \cite{abramoff:IEEERBE:2010}, and the segmentation of fluid using the active contours approach by \cite{fernandez:IEEETMI2005}.

Other approaches used for the segmentation of retinal OCT include: The traditional graph-cut methods by \cite{Salazar:jbhi2014,Salazar:icarcv2010,Salazar:cimi2011,Kaba:oe2015}, the Markov Random Fields by \cite{Salazar:his2012,Wang:jbhi2017}, probabilistic modelling \cite{Kaba:hiss2014,Kaba:his2013}, dynamic programming by \cite{Chiu:BOE2015}, level set by \cite{Dodo:access2019,Dodo:bioimaging2019} and a combination of a fuzzy C-means and level set contour by \cite{wang:BOE2016}.

Recent approaches have shifted to the Deep Learning methods, some of which will be reviewed briefly below.

\cite{Fang:BOE2017} presented an approach to segment nine retinal boundaries from retinal OCT using a combination of convolutional neural network (CNN) and graph search. Their approach was tested on 60 volumes consisting of 2915 B-scans from 20 human eyes suffering with dry AMD. 
\cite{Fauw:NatureMedicine2018} presented a 3D U-Net \cite{Cicek:MICCAI2016} model framework for diagnosis and referral in retinal disease. Their dataset consisted of 14,884 volumes from 7,621 patients.
The ReLayNet was presented by \cite{Roy:BOE2017}, which is a 2D-like U-Net architecture to segment layers and fluids in the OCT images. The 
method was validated on the Duke DME dataset \cite{Chiu:BOE2015} which consists of 110 B-Scans from 10 patients suffering from Diabetic Macular Edema (DME).
Another CNN approach was reported by \cite{Lee:BOE2017} to segment fluid from 1,289 OCT images from patients suffering from Macular Edema (ME). 
\cite{Lu:arXiv:2017} reported a CNN approach to detect and segment three retinal fluid types from OCT images and their method was validated on the RETOUCH dataset \cite{bogunovic:IEEETMI2019}. 
A neutrosophic transformation and a graph-based shortest path to segment fluid in OCT images was presented in \cite{Rashno:IEEETBE2017}. Their method  was also evaluated on the DME dataset.
A Deep Learning approach for simultaneous segmentation of layers and fluids in retinal OCT B-Scans from patients suffering from AMD is proposed in \cite{ndipenoch:IEEE2022}. The algorithm consists of the traditional U-Net with an encoding and a decoding path, skip connection blocks, squeeze and exiting blocks and an Atrous Spatial Pyramid Pooling (ASPP) block. The method is validated on 1136 B-Scans from 24 patients.
Other CNN approaches to segment fluids in retinal OCT modality includes
\cite{schlegl:ICIPMI2015,venhuizen:BIO2018,Gopinath:IEEEJBHI2018,Girish:IEEEJBHI2018}.

Previous studies indicate that in this domain U-Net and CNN are the most popular methods used but, U-Net tends to outperform CNN, and hence U-Net is the preferred choice in many applications. An overview of the related work is summarised in Table~\ref{tab:revew} below.
%\todo[inline]{discuss Table 1 in the text.}

\begin{table}[H]
\addtolength{\tabcolsep}{-6pt}

\caption{Overview of the related work with references and corresponding fluid and disease types.}
\label{tab:revew}
\centering
\begin{tabular}{@{}lllllll@{}}
\toprule
Reference  & Class  &  Disease \\ 
  \midrule

\cite{fernandez:IEEETMI2005}   & Fluid  & AMD \\

\cite{garvin:IEEETMI2009}   & Fluid      & --   \\ 

\cite{abramoff:IEEERBE:2010}  & Fluid     & ME  \\

\cite{Salazar:icarcv2010}   & --  & --         \\

\cite{Salazar:his2012}  & Optic disc   & DR   \\

\cite{Salazar:jbhi2014}   & B. vessel     &  DR  \\

\cite{Chiu:BOE2015}  & Fluid   & DME    \\

\cite{wang:BOE2016}  & Fluid   & DME   \\

\cite{Wang:jbhi2017}  & Fluid   & DME   \\

\cite{Fang:BOE2017}  & Layers   & AMD   \\

\cite{Dodo:access2019}  & Layers   & --   \\

\cite{loo:BOE2018}   & Fluid   & ME   \\

\cite{schlegl:ICIPMI2015}  & Fluid   & --  \\

\cite{venhuizen:BIO2018}   & Fluid   & AMD  \\

\cite{Gopinath:IEEEJBHI2018}   & Fluid   & DME  \\

\cite{Girish:IEEEJBHI2018}  & Fluid   & ME  \\

\cite{Lu:arXiv:2017}  & Fluid   & ME  \\

\cite{Fauw:NatureMedicine2018}  & Fluid   & --  \\

\cite{Roy:BOE2017}  & Fluid   & DME   \\

\cite{Rashno:IEEETBE2017}  & Fluid   & DME  \\

\cite{ndipenoch:IEEE2022}  & Fluid/Layer   & AMD  \\

\end{tabular}
\end{table}

\section{\uppercase{Method}}
\label{sec:methods}

% \subsubsection{Problem Statement}
% \label{subsuc:problem_statement}

Deep Learning methods have had success in image segmentation (pixel-wise classification) but this depends hugely on large datasets. In medical imaging obtaining a dataset is very challenging and often very small and limited. We aim to provide a model that performs very well on very small and limited datasets of less than a hundred training images. 
% In this work for every input image we aim to classify each pixel correctly to belong to one of the ten classes. 

% \subsection{CoNet}

\begin{figure*}
% \centerline{\includegraphics[width=5cm \textwidth]{resplus.png}}
\centerline{\includegraphics[width=0.8\textwidth ] {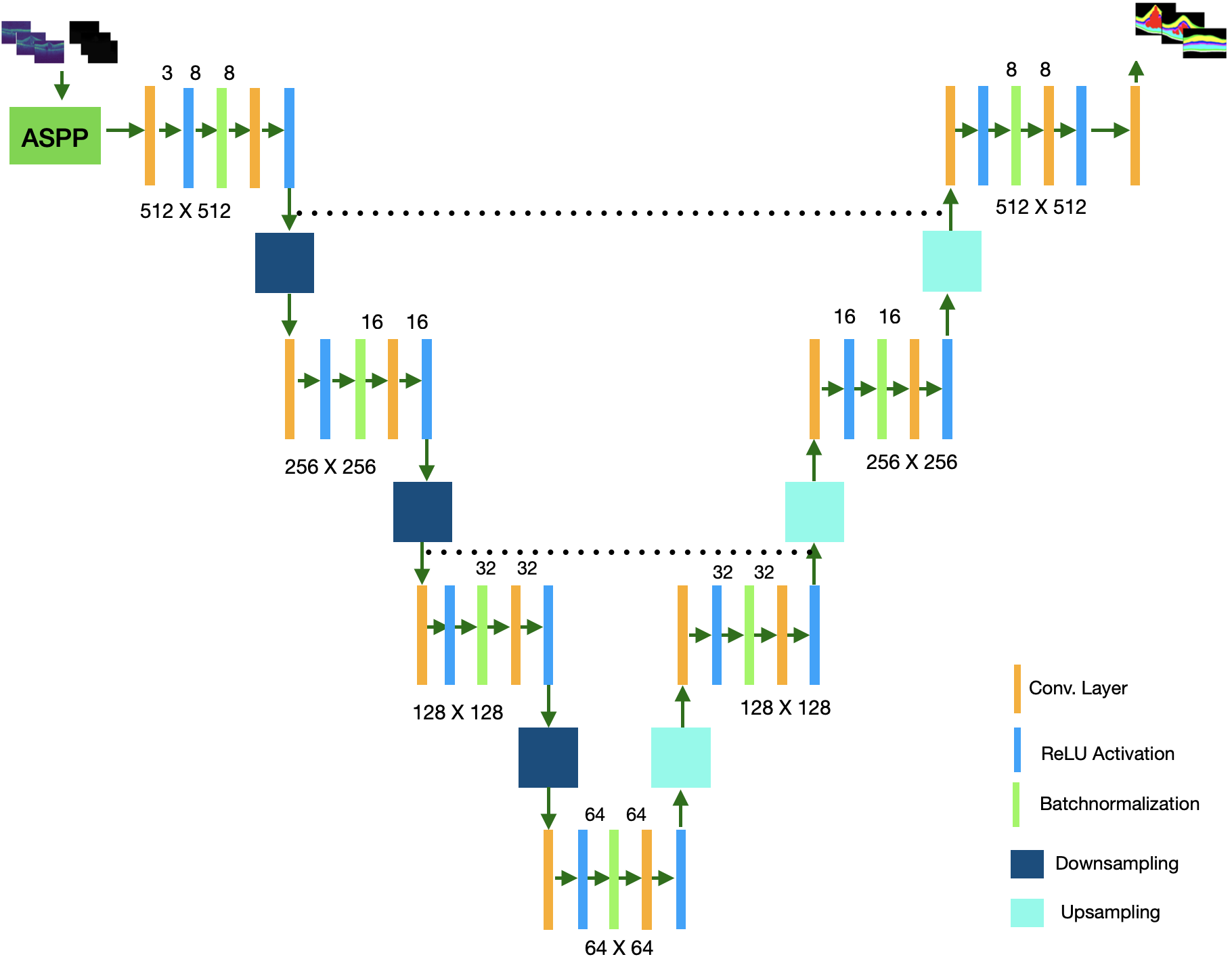}}
% \caption{A high level structure of the CoNet demonstrating the encoding and decoding paths, the bottleneck, the classification layer and the ASPP block.}
\caption{Architecture of the proposed CoNet, consisting of the ASPP block, the encoding path, the bottleneck, the decoding path and the classification layer.}
\label{fig:CoNet_archiecture}
\end{figure*}

The proposed CoNet model is based on commonly used  U-Net architecture \cite{Ronneberger:MICCAI2015} but adapted to the specific problem of retina image segmentation on very small datasets. The model architecture consists of an encoding path, a decoding path, a bottleneck, a classification layer, and an Atrous Spatial Pyramid Pooling (ASPP) block, as shown in Figure~\ref{fig:CoNet_archiecture}. 
In this section, we will explain our method, the changes we made and how it is different to the U-Net.

\subsection{Encoding Path}
\label{subsuc:encoding_path}
The encoding path is use to capture local contextual and spatial information. As we move down the encoding or contracting path the feature map is reduced by half after every convolutional block by a convolutional operation at the downsampling layer. A total of three convolutional blocks are used and for each block the convolutional operations are set up in the order of: (1) convolutional layer which converts all the pixels of the receptive field into a single value and passes it to the next operation, (2) ReLU activation, to circumvent  the problem of vanishing gradient, (3) batch normalisation layer to prevent over-fitting during training, (4) convolutional layer, and (5) ReLU activation. For the convolutional operations a rectangular kernel size of $9\times3$ is used to match the rectangular shape of the original B-Scans as opposed to the square kernel size of $3\times3$ used in U-Net, also to ensure that the feature map before and after the convolutional layer is the same padding was set to $3\times1$ still to match the rectangular shape of the original B-Scans and finally to ensure no overlapping when constructing the feature map a stride of 1 was used.
%\todo[inline]{replace all x in 3x3 the likes by $\times$}

\subsection{Decoding Path}
\label{subsuc:decoding_path}
The decoding path is used to enable precise localization of the pixel and as we move up the decoding or expansive path, before each convolutional block the size of the feature maps is double  by a convolutional operation at the upsmapling layer. Same as in the encoding path a total of three convolutional blocks are used and set up in the same order as mentioned in \ref{subsuc:encoding_path}. In addition to that, the upsampling layer was used to double the size of the feature map by capturing spatial information from the previous feature map and also to ensure that the size of the input image is the same as the output image, the concatenating layer is used  to concatenate images from the encoder phase to their corresponding decoder phase.

Because of the very small size of the dataset (only 55 B-Scans for training), we reduce the depth of the network from 5 convolutional blocks as in the standard 2D U-Net to 3 in both the encoding and decoding phase. Furthermore reducing the depth of the network trains the model faster and uses less memory because less parameters are used.

\subsection{Bottleneck}
\label{subsuc:bottleneck}
Between the encoding and decoding paths is a bottleneck. The bottleneck serves as a bridge layer between the encoding or contracting path and the decoding or expansive path to ensure a smooth transition from one path to the other.
In CoNet, the bottleneck is made up of a convolutional block that consists of six parts or layers in the order of  convolutional layer, ReLU activation, batch normalization layer, convolutional layer, and ReLU activation. These layers were used for the same reasons as mentioned in section \ref{subsuc:encoding_path}

\subsection{Atrous Spatial Pyramid Pooling (ASPP)}
\label{subsuc:aspp}
The ASSP is a technique used to capture global contextual information on a multi-scale by applying multiple parallel filters with different frequencies or dilating rates on a given image or feature map \cite{chenIEEETPMAI:2017}.
To enhance the performance of the ASSP block, global average pooling is used at the last feature map to further capture global information. The output of the parallel filters are concatenated using a $1\times1$ convolution to get the final results. While ASSP is designed to capture global information, it is also computational efficient.

\begin{figure*}
\centerline{\includegraphics[width=.8\textwidth]{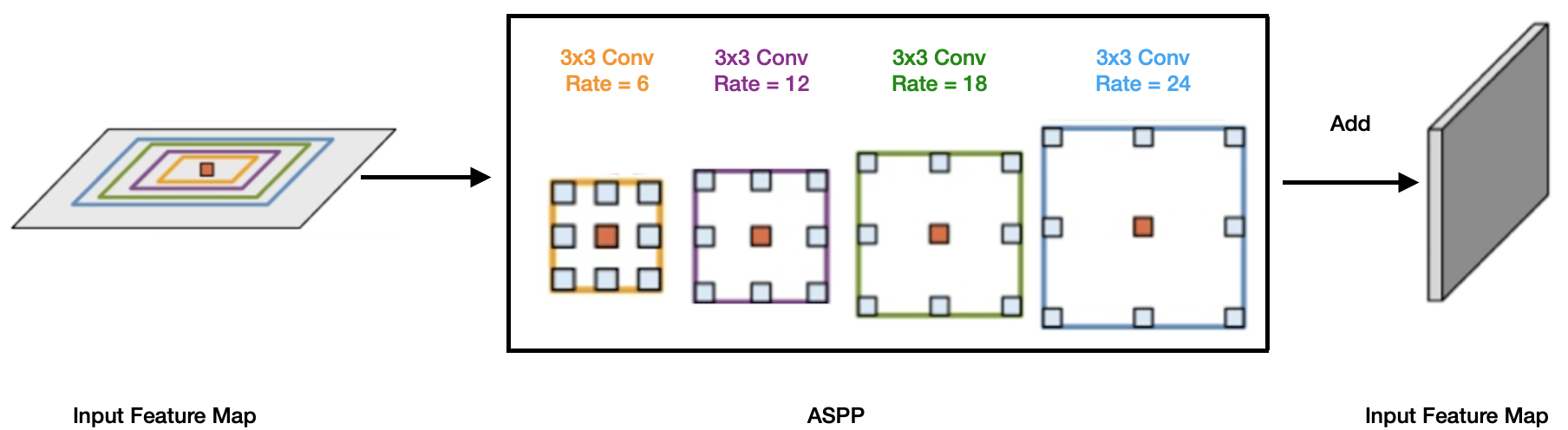}}
% \centerline{\includegraphics[width=6cm \textwidth]{ASPP.png}}
% \caption{An exmaple of capturing multi-scale information by applying multiple parallel filters with different frequencies using an ASSP block.}
\caption{The ASPP block in the proposed model captures multi-scale information by applying multiple parallel filters with different frequencies.}
\label{fig:ASPP_Structure}
\end{figure*}

No ASPP block is used in the standard 2D U-Net. We have used an ASPP block as the input layer of CoNet, and it consists of 4 parallel filters with a dilating rate of 6, 12, 18, and 24. To circumvent the problem of high fluid variability (the fluid class was absent in some B-Scans for some patients) and an imbalance dataset we have used the ASPP block in CoNet.
% The diagram of the ASPP block used in CoNet is shown in figure 
The ASPP block used in CoNet is illustrated in Figure~\ref{fig:ASPP_Structure}.

\subsection{Classification Layer}
\label{subsubsection:dense_layer}
At the classification layer, we have used a convolutional layer with a kernel size of 3, stride of 1 and padding of 1. The task is to determine which class out of the ten labelled classes each voxel or pixel of the final feature map is assigned to. In the 2D U-Net the same goal was achieved using the SoftMax activation.

\section{\uppercase{Experiments}}
\label{sec:experiments}
\subsection{Dataset}
\label{subsubsection:dataset}

\begin{figure*}
\centerline{\includegraphics[width=.8\textwidth]{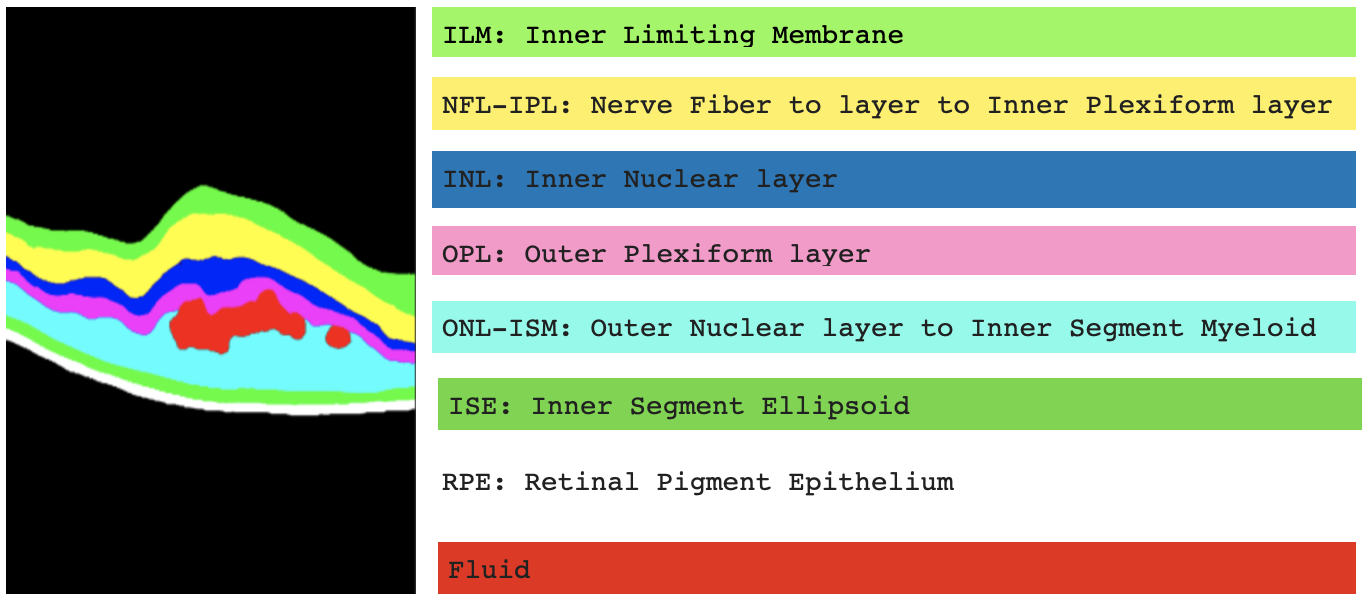}}
% \caption{A demonstration of annotation and labelling of the 10 classes (7 layers, 2 backgrounds and 1 fluid) in the Duke DME dataset. 
% }
\caption{Annotation and labelling of the 10 segments (7 retinal layers, 2 backgrounds and 1 fluid) in the Duke DME dataset. 
}
\label{fig:Duke_Datatset_annotation}
\end{figure*}

The dataset used in the experiments is the Duke DME dataset \cite{Chiu:BOE2015} which is publicly available. It is consists of 110 B-scans from 10 patients with severe DME pathology. It was collected using the standard Spectralis (Heidelberg Engineering, Heidelberg, Germany). The volumetric scans were Q = 61 B-scans N = 768 A-scans with an axial resolution of 3.87 µm/pixel, lateral resolution ranging from 11.07 - 11.59 µm/pixel, and azimuthal resolution ranging from 118 - 128 µm/pixel. 
Annotation of the images was done by 2 human experts for three categories (layer, fluid and background) consisting of 10 classes (1 fluid, 2 backgrounds and 7 layers). 
In the past the retinal OCT is layered for 10 layers but for lucidity they are grouped into 7 distinct classes which are: Inner Limiting Membrane (ILM), Nerve Fiber ending to Inner Plexiform Layer (NFL-IPL), Inner Nuclear Layer (INL), and the Outer plexiform Layer (OPL).
A fluid class was identified and the two background classes were the area above the retinal and the area below the retinal.

In this work the annotated colors for the classes are: Black which is the area above and below the retinal, light green which is the ILM layer, yellow which is the area between NFL and IPL layers, Blue which is the INL, Pink which is the OPL layer, light blue which is the area between the ONL and ISM layers, Green which is the ISE layer, White which is the RPE, and Red which is the fluid.
An example of annotation and labelling of classes is shown in Figure~\ref{fig:Duke_Datatset_annotation}.

It is worth to note that the Duke DME dataset was collected for two problems (layer and fluid segmentation). Also to add to the complexity of the dataset, the fluid class demonstrates a high level of variability and was not present in some B-Scans for some patients.

\subsection{Training and Testing}
\label{subsubsection:trainin_testing}
Retinal OCT layers are complex in nature, coupling with high level of variability of the fluid classes. It is therefor a common practice to do segmentation of layers and detection of fluids separately, but in this work we performed both simultaneously which is a harder task.

In this work training and testing were done using annotation from expert 2.

Training was done on 55 B-Scans and no data augmentation was used. We used B-Scans instead of the entire volumes because of the anisotropic resolution of OCT volumes and the present of possible motion artifacts across B-scans.

K-fold cross validation was used for training, validation and testing. Parameters and environmental settings where the same for the proposed model and the comparison models to ensure fairness.
B-Scans from 5 patients were used per fold, that is patients 1-5 in the first fold and 6-10 in the second. To eliminate bias the use of adjacent B-Scans in training, validation, and testing is not recommended.
Across all the experiments the parameters were set up as, still, the same as in the comparison models: 
the value of k was 2, the original B-Scans were resized to $512\times512$ pixels, the loss function used was Categorical cross-entropy which provides an estimated probability between the predicted voxels and the ground truth for the current state of the model, the batch size was set to 4, the cost function was optimized using AdaDelta and back-propagation using the chain rule, by default the learning rate is set by AdaDelta as explained in \cite{zeiler:preprint2012}, and the model was trained for 200 epochs. The AdaDelta's equation is shown in Eqn~(\ref{eqn:adad}).

\begin{equation}
\label{eqn:adad}
%   \caption{AdaDelta Formula}
    \Delta \theta_t = - {\frac{ n } {\sqrt{E[g^2]_t + \epsilon}}}g_t
\end{equation}

Dice score also known as the F1 Score was the evaluation metric used to measure the performance of the algorithm. It gives a score of how well the pixels are classified to belong to the correct class per class in the range from 0 to 1 with 0 being the worst and 1 the perfect classification. In many medical image segmentation problems Dice Score is the preferred choice. The formula to calculate the Dice score is shown in Eqn~(\ref{eqn:dsc}).

% \begin{equation}
% \label{eqn:dsc}
% \vcenter{\hbox{\begin{minipage}{2cm}
% \centering
% \includegraphics[width=2cm,height=2cm]{DSC.png}
% %\captionof{figure}{Dice Score calculation.}
% \end{minipage}}}
% \qquad\qquad
% \begin{aligned}
% DSC = \frac{ 2 |X \cap Y| } {|X|+|Y|}
% \end{aligned}
% \end{equation}

\begin{equation}
\label{eqn:dsc}
DSC = \frac{ 2 |X \cap Y| } {|X|+|Y|}
\end{equation}
 
The fluid class was missing for some B-Scans for some patients. Hence during testing the calculation of Dice score for the fluid class was exempted from B-scans with no fluid reference for that patient to avoid over estimation or under estimation.

The models were trained on a GPU work station with NVIDIA RTX A6000 48GB. 
The models were implemented in Python, using PyTorch library.

\subsection{Results}
\label{subsubsection:results}
In this section we present and analyse the segment class results measured in Dice score, of the proposed CoNet and compare them to the comparison models (the state-of-the-art, ReLayNet and baselines U-Net) and the human expert annotation (Inter-observer). 

\begin{figure*}
\centering{\includegraphics[width=15cm ]{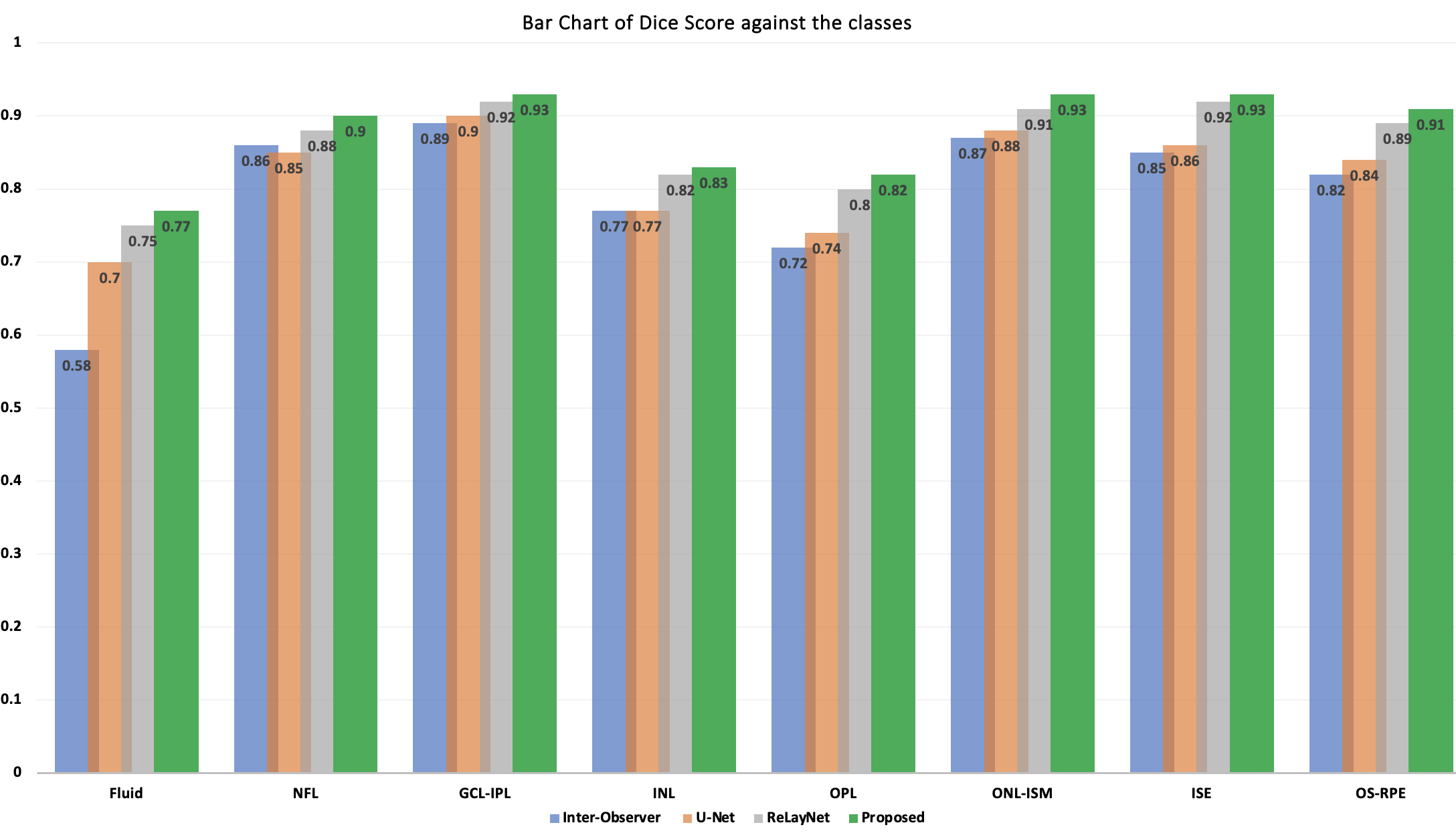}}
\caption{A Bar chart comparison of the performance in Dice score grouped by segment class of the inter-observers, U-Net, RelayNet and the proposed CoNet. 
}
%\vspace*{0mm}
\label{fig:bar_chart}
\end{figure*}

\begin{figure*}
\centering
% \centerline{\includegraphics[width=10cm \textwidth]{Result_summary.png}}
\centerline{\includegraphics[width=15cm ]{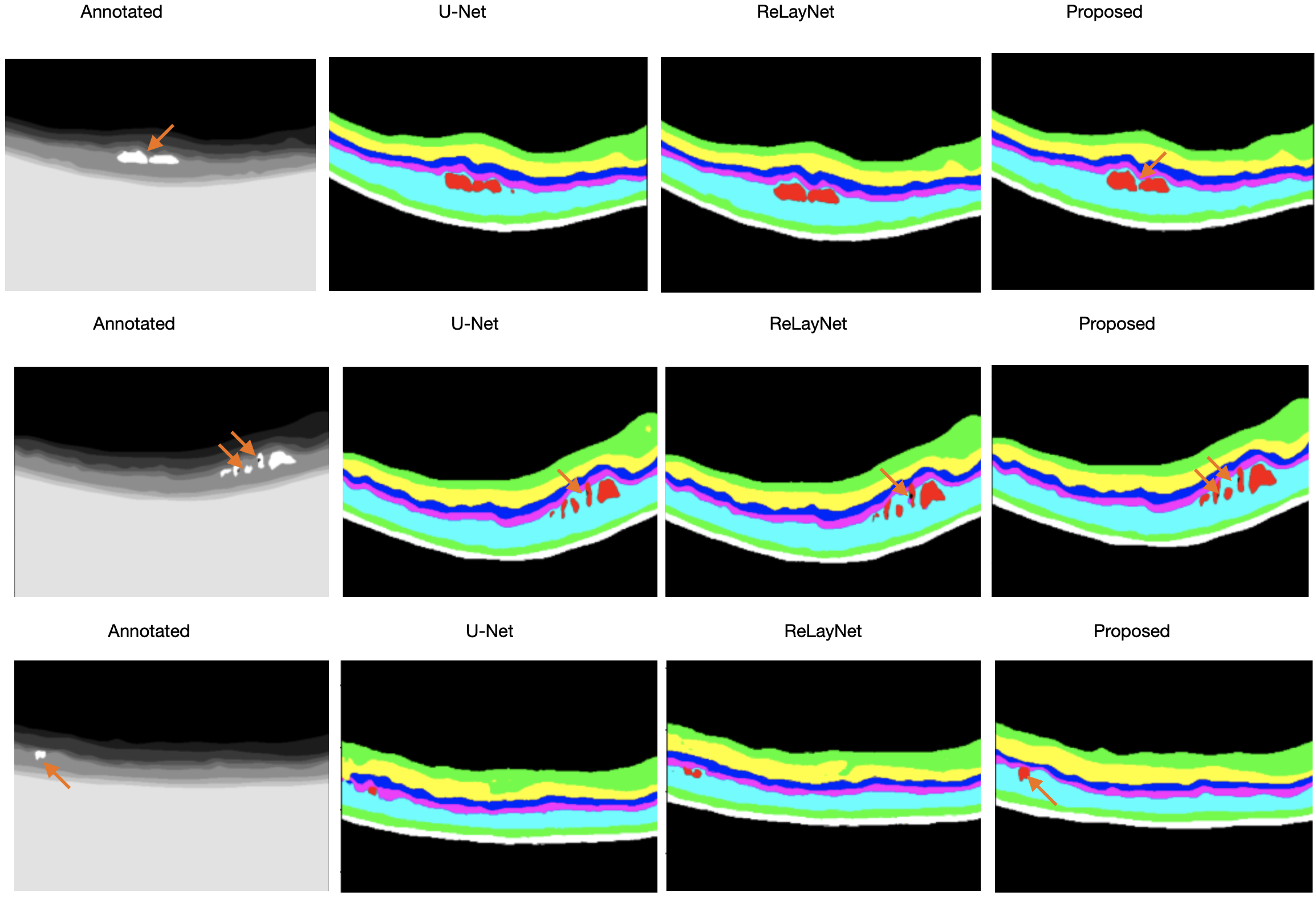}}

\caption{Examples to illustrate the visualisation output of U-Net, ReLayNet and the proposed CoNet, in order of the inputs, annotations and outputs with orange arrows to demonstrate fine details picked up by the models. }
%\vspace*{-30mm}
\label{fig:result_visualization}
\end{figure*}

\begin{figure*}[ht]
\centering
\centerline{\includegraphics[width=16cm]{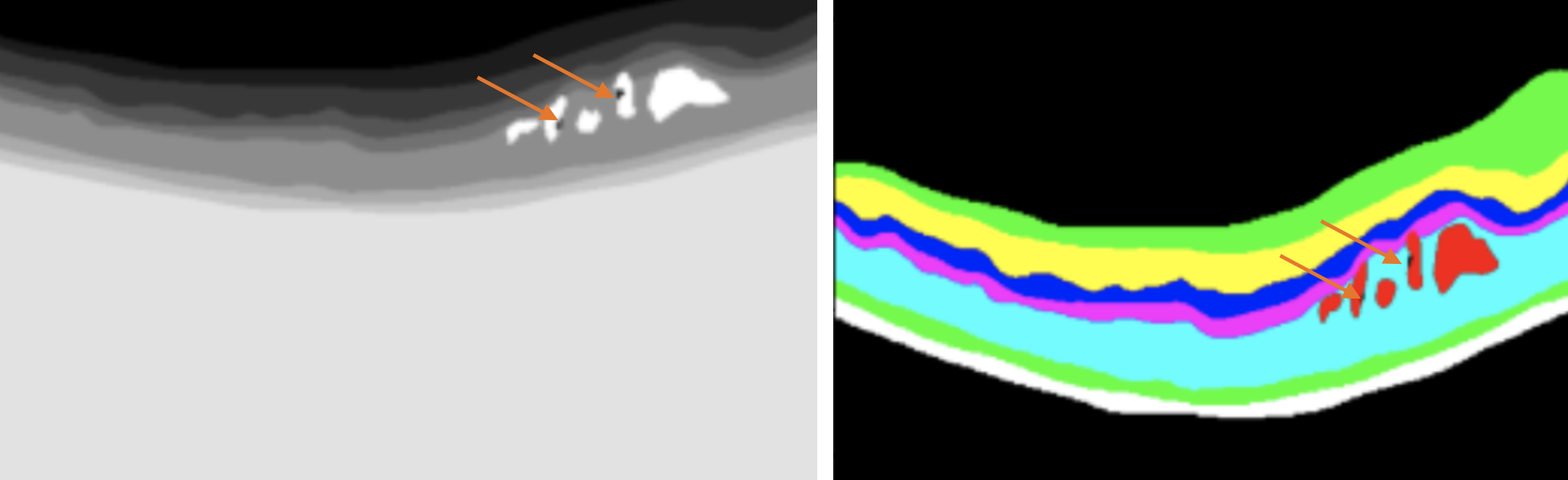}}

\caption{A Zoom in of the B-scan of the second row in Figure~\ref{fig:result_visualization} to highlight the fine details picked up by the CoNet using orange arrows.}
\label{fig:zoom_in}
\end{figure*}

A bar chart of the segmentation grouped by segment classes is shown in Figure~ \ref{fig:bar_chart}, the Dice Scores in Table~\ref{tab:results_table}, examples of the visualization results together with their annotations are illustrated in Figure~ \ref{fig:result_visualization} and a zoom in of a visualization output example from CoNet is illustrated in Figure~ \ref{fig:zoom_in}. Orange arrows are used to show fine details in the annotated B-Scans that were picked up by corresponding models. Analysis from our results show that: 

    \begin{enumerate}
    \item The proposed model CoNet outperforms the human experts, the  baseline (U-Net) and current state-of-the-art model ReLayNet in every single class by a clear margin.
    \item CoNet obtained a Dice Score of 77\% which is 19\% greater than the human experts' (inter-observer) in the fluid class which was the most difficult to segment.
    \item We obtained a Dice score of 90\% and above in 8 out of the 10 classes.
    \item The baseline U-Net, the state-of-the-art architecture RelayNet, and the proposed CoNet all obtained a perfect Dice Score of 100\% in both background classes (area above and below the retinal).
    \item CoNet obtained an overall mean Dice Score of 88\% which is 8\% higher than that of the human experts' annotation results of 80\%.
    \item We noticed an increase of performance from the standard U-Net to a shallower and less complex architectures in the order of ReLayNet, and CoNet.
    
\end{enumerate}

\begin{table}
\addtolength{\tabcolsep}{-2pt}
% \caption{Table of the Dice Score per class per model and Inter-observer }
% \caption{A table of summary of Dice Scores by segment classes (rows) and models (columns). }
\caption{Segmentation performance (Dice Scores) by segment classes (rows) and models (columns). }
\label{tab:results_table}
\centering
\begin{tabular}{@{}lllllll@{}}
\toprule
%  & Inter-observer & Baseline UNet & Proposed UNet & Proposed ResUNet & Proposed ResUNet\texttt{++} \\
 & Inter\_Obs.  & U-Net  &  ReLayNet & \thead{Proposed}\\ 
  \midrule
 
Fluid      & 0.58     & 0.70     & 0.75   & \textbf{0.77}  \\ 

NFL        & 0.86     & 0.85      & 0.88   & \textbf{0.90}
               \\
GCL\_IPL   & 0.89     & 0.90      & 0.92   & \textbf{0.93}   \\

INL        & 0.77     & 0.77      & 0.82   & \textbf{0.83}  \\

OPL        & 0.72     & 0.74      & 0.80   & \textbf{0.82}   \\

ONL\_ISM   & 0.87     & 0.88      & 0.91   & \textbf{0.93}    \\

ISE        & 0.85      & 0.86     & 0.92   & \textbf{0.93}   \\

OS\_RPE     & 0.82     & 0.84     & 0.89   & \textbf{0.91}               

\end{tabular}
\end{table}

\section{\uppercase{Conclusions}}
\label{sec:conclusion}

In this paper, we have presented the CoNet, a Deep Learning approach 
% adapted from previous studies by \cite{Ronneberger:MICCAI2015} 
for the joint layers and fluids segmentation of retinal OCT B-Scans. The model was evaluated on the publicly available Duke DME dataset consisting of 110 B-Scans without any data augmentation.  Taking into consideration of the specific characteristics of the problem, in particular with small available dataset, the proposed model has the following distinct features compared to the previous U-Net based models:
% we have adapted the \cite{Ronneberger:MICCAI2015} model as follows: 

\begin{enumerate}
\item 
We have reduced the depth of the network from 5 to 3 convolutional blocks. This was done because of the very small size of the dataset (only 55 B-Scans were used during training and the rest 55 for testing). Deeper and more complex architectures turn to yield poorer results. Also, reducing the depth of the network enhances the training speed of the network and uses less memory since less parameters are used. 
\item 
We have introduced an ASPP block at the input layer to capture global information from the input image, because the dataset demonstrates a high level of variability. The fluid class was not present in some B-Scans for some patients. 
\item
At the classification layer to classify each pixel to belong to one of the 10 classes we have used a convolutional layer instead of the SoftMax activation. This was because using the convolutional layer for the classification of the fluid class which is highly variable yielded a better and more accurate results. 
\item
We have used a rectangular kernel size of $9\times3$ instead of the  square kernel size of $3\times3$ to match the rectangular shape of the original B-Scans. 
\end{enumerate}

Evaluation was done on the basis of Dice Score which is a standard method of evaluating segmentation problems. Experimental results show that the proposed model outperformed both the human experts' annotation and the current state-of-the-art architectures by a clear margin, even on a very small, imbalanced and complex dataset with a high degree of presence of pathology that severely affects the normal morphology of the retina.

The dataset was collected for 2 problems (layers and fluid segmentation) which can be experimented separately but we decided to do both jointly together which is a more challenging task. 
  
The CoNet can be directly applied to solve real world problems and to monitor the progress of eye diseases such as diabetic macular edema (DME), age-related macular degeneration (AMD) and Glaucoma. In the future we will evaluate the CoNet on other benchmark datasets, and compare our results to other state-of-the-art models. Also we plan to extend the current 2D  network to 3D. 

\bibliographystyle{apalike}
{\small
%\clearpage
\bibliography{example}}

\end{document}